%% file: main.tex
\documentclass[11pt,a4paper]{article}
\usepackage{resemble-paper}

\title{DETECT-3B-Omni is Agnostic\\of Content and Demographics}
\shorttitle{DETECT-3B-Omni is Agnostic of Content and Demographics}
\venue{Resemble.ai \& DT}
\headerlogos{figures/resemble-logo,figures/telekom-logo}

\author{%
  \textbf{Nicolas M.\ M\"uller}\affmark{1},\quad
  \textbf{Aditya Tirumala Bukkapatnam}\affmark{1},\quad
  \textbf{Dominik Schnieders}\affmark{2},\quad
  \textbf{Zohaib Ahmed}\affmark{1}
  \\[4pt]
  \affiliation{1}{Resemble AI, Mountain View, CA, USA}\quad
  \affiliation{2}{Deutsche Telekom, Bonn, Germany}
  \\[2pt]
  \email{\{nicolas,aditya,zohaib\}@resemble.ai}\quad
  \email{Dominik.Schnieders@telekom.de}
}

\date{}

\begin{document}
\maketitle

\begin{abstract}
A trustworthy and GDPR-compliant deepfake audio detector must base its
decisions on acoustic artifacts, not on what is being said or who is
speaking.
We present a large-scale study of \emph{semantic independence} for
Resemble AI's detector, DETECT-3B-Omni.
Using 10,240 audio samples from diverse US English speakers across 30
states, generated through 8 different AI voice-cloning systems, we test
whether detection accuracy depends on spoken content (benign versus
malicious), speaker gender, speaker age, or speaker region.
Using equivalence testing, our results show that the accuracy
difference between any two of these groups is at most 2 percentage
points, at 99\% confidence.
The detector therefore identifies AI-generated audio with equivalent
accuracy regardless of what the audio says or who the speaker is.
\end{abstract}

\section{Introduction}
\label{sec:intro}

Audio deepfake detection systems are increasingly deployed to protect
phone calls and voice authentication against AI-generated fraud.
A key use case is real-time monitoring of telephone calls to flag AI-generated speech before it can cause harm.
Under the EU General Data Protection Regulation (GDPR) and applicable telecommunication law, deploying such
a system requires demonstrating that it processes only the information
strictly necessary for its purpose: detecting whether audio is
AI-generated.
In particular, the detector must not make decisions based on the
\emph{content} of the speech (what is said) or the \emph{identity} of
the speaker (who is speaking), as this would constitute unnecessary
processing of personal data.

A detector that is \emph{semantically independent} satisfies this
requirement: its output depends only on acoustic artifacts that
distinguish human speech from AI-generated speech, not on the words
being spoken or the characteristics of the speaker.
If a detector's accuracy changed depending on the topic of
conversation (for example, performing differently on financial
discussions versus casual speech), it would effectively be ``listening
to'' the content, raising both fairness and GDPR compliance concerns.
Similarly, if detection accuracy varied across speaker demographics
(gender, age, or regional accent), the system would be unreliable for
certain populations and potentially discriminatory.

We present a controlled, large-scale study to verify that DETECT-3B-Omni
exhibits no such dependence.
We construct a dataset of 10,240 audio samples covering:
\begin{itemize}\itemsep0pt
  \item 640 unique sentences, split evenly between benign (everyday
    business communication) and malicious (social engineering) content,
  \item recordings by diverse native US English speakers, balanced
    across gender and spanning ages 20--55 and 30 US states,
  \item fake audio generated by 8 state-of-the-art open-source
    text-to-speech voice-cloning models.
\end{itemize}

We classify every sample using the DETECT-3B-Omni API and
compare detection accuracy across content categories and demographic
groups.
The goal is straightforward: if the detector is semantically independent,
accuracy should be the same regardless of how we split the data.
This provides evidence that the system is suitable for deployment in
GDPR-regulated environments, where it can monitor calls for deepfake
audio without processing the semantic content of conversations.

\section{Data}
\label{sec:data}

\subsection{Sentence Design}

We designed 640 English sentences in two categories of 320 each:

\begin{itemize}\itemsep0pt
  \item \textbf{Benign}: standard business and technical communication
    (e.g., invoice confirmations, meeting scheduling, account inquiries).
  \item \textbf{Malicious}: social engineering and fraud scenarios
    (e.g., urgent payment demands, credential phishing, authority
    impersonation).
\end{itemize}

Sentences are of comparable length (115--150 characters, mean 131) to
ensure similar audio duration.
Sentences were generated using a large language model and manually
reviewed to confirm category membership and linguistic diversity.
\tabref{tab:examples} shows an example pair from the IT/security domain.
The full list of 80 categories is provided in Appendix~\ref{app:categories}.

\begin{table}[h]
\centering
\small
\begin{tabular}{lp{0.78\textwidth}}
\toprule
\textbf{Label} & \textbf{Sentence} \\
\midrule
Benign &
\emph{``I just popped into the local carrier store to swap out my old
SIM card for a 5G version and finally boost my data speeds.''} \\[4pt]
Malicious &
\emph{``Yo, just a quick heads up that your SIM card needs a safety
boost. Text back that PAC code real quick so you don't lose service
tonight.''} \\
\bottomrule
\end{tabular}
\caption{Example sentence pair from the mobile/telecom domain.}
\label{tab:examples}
\end{table}

\subsection{Real Audio}

Each of the 640 sentences was recorded by 8 native US English speakers,
yielding 5,120 bona-fide recordings.
Speakers are balanced across gender (male and female), span ages 20--55,
and represent 30 US states.
All recordings are 48\,kHz, 16-bit stereo WAV files.

\subsection{Fake Audio}

We generated fake (spoofed) versions of the same sentences using 8
open-source voice-cloning TTS models:

\begin{enumerate}\itemsep0pt
  \item Chatterbox~\cite{chatterbox}
  \item Chatterbox-Turbo~\cite{chatterbox}
  \item Qwen3-TTS 1.7B~\cite{qwen3tts}
  \item Qwen3-TTS 0.6B~\cite{qwen3tts}
  \item XTTS2~\cite{xtts}
  \item E2-TTS~\cite{e2tts}
  \item F5-TTS~\cite{f5tts}
  \item MegaTTS3~\cite{megatts3}
\end{enumerate}

Each sentence is assigned to one model (round-robin), and synthesized
8 times (once per reference speaker) using a recording of that speaker
from a different sentence as the voice reference.
This produces 5,120 spoofed recordings, matching the real
audio count.

\input{results/dataset}

\section{Experiment Design}
\label{sec:experiments}

Every audio sample is classified by the DETECT-3B-Omni API,
which returns a prediction of \emph{real} or \emph{fake}.
We store the full API response for each sample as evidence.

\subsection{Notation}

We partition the dataset into four groups based on two dimensions:

\begin{itemize}\itemsep0pt
  \item \textbf{Content}: benign (B) or malicious (M)
  \item \textbf{Audio}: real (R) or fake (F)
\end{itemize}

\noindent This gives four atomic groups:

\begin{table}[h]
\centering
\begin{tabular}{lcc}
\toprule
 & \textbf{Real (R)} & \textbf{Fake (F)} \\
\midrule
\textbf{Benign (B)} & BR & BF \\
\textbf{Malicious (M)} & MR & MF \\
\bottomrule
\end{tabular}
\caption{Four atomic groups. Each contains 2,560 samples.}
\label{tab:groups}
\end{table}

Every experiment below compares two subsets of data and tests whether
detection accuracy differs between them.

\subsection{Content Independence}

These experiments test whether the \emph{meaning} of the spoken text
affects detection accuracy.

\begin{itemize}\itemsep0pt
  \item \textbf{E1: All audio, benign vs.\ malicious.}
    Compare $\text{BR} \cup \text{BF}$ against
    $\text{MR} \cup \text{MF}$.
    If the detector is content-independent, accuracy should be the same
    for benign and malicious sentences.
  \item \textbf{E2: Fake audio only.}
    Compare BF against MF.
    Tests whether the detector is more or less likely to catch
    AI-generated audio depending on what it says.
  \item \textbf{E3: Real audio only.}
    Compare BR against MR.
    Tests whether the detector is more or less likely to trust human
    speech depending on what it says.
  \item \textbf{E4: Maximum contrast.}
    Compare $\text{BF} \cup \text{MR}$ (benign-fake + malicious-real)
    against $\text{BR} \cup \text{MF}$ (benign-real + malicious-fake).
    This deliberately pairs content with opposing labels to maximize
    any potential content bias.
\end{itemize}

\subsection{Demographic Independence}

These experiments test whether \emph{speaker characteristics} affect
detection accuracy.

\begin{itemize}\itemsep0pt
  \item \textbf{E5: Gender.}
    Compare detection accuracy for male versus female speakers.
  \item \textbf{E6: Age.}
    Compare speakers under 40 versus speakers 40 and older.
  \item \textbf{E7: Region.}
    Compare speakers from east of the Mississippi River (20 states,
    covering the Eastern Seaboard and Midwest) against speakers from
    west of the Mississippi (10 states, covering the South and
    West Coast).
\end{itemize}

\subsection{Evaluation: Equivalence Testing}

For each experiment, we do not merely check whether a difference
\emph{exists} between two groups. Instead, we show that the difference
is \emph{negligibly small}.
Concretely, we show that the accuracy difference between any two groups
falls within $\pm 2$ percentage points, with 99\% confidence.

We compute a 99\% confidence interval (CI) for the difference in
accuracy between the two groups.
If the entire CI lies within the equivalence margin
$[-2\,\text{pp},\;+2\,\text{pp}]$, the groups are declared
\emph{equivalent}, meaning we have asserted with high confidence
that the detector treats both groups the same.

Formally, let $\hat{p}_A$ and $\hat{p}_B$ be the observed accuracy
rates for groups $A$ and $B$ with $n_A$ and $n_B$ samples respectively.
The 99\% CI for $\Delta = \hat{p}_A - \hat{p}_B$ is:
\begin{equation}
  \Delta \;\pm\; 2.576 \cdot
  \sqrt{\frac{\hat{p}_A(1-\hat{p}_A)}{n_A}
       +\frac{\hat{p}_B(1-\hat{p}_B)}{n_B}}
\end{equation}
where 2.576 is the 99.5th percentile of the standard normal distribution
(the standard critical value for a two-sided 99\% confidence interval).
If $\text{CI} \subset [-2\,\text{pp},\;+2\,\text{pp}]$, the groups are
equivalent.

The $\pm 2$\,pp margin is non-trivial at the per-group sample sizes
used in this study: as shown in
Appendix~\ref{app:random-subsets} (\tabref{tab:random_subsets}), a
purely random split of a smaller atomic subset ($n=1{,}280$ per half)
already produces CIs whose width can exceed $\pm 2$\,pp solely from
sampling noise.
Passing the equivalence test at this margin therefore reflects a
genuine absence of group-dependent effects, not a trivially wide
tolerance.

\subsection{Comparison to a Clinical Non-Inferiority Trial}
\label{sec:comparison-capit}

To put our choice of margin and confidence level in context, we
compare against the CAP-IT trial~\cite{bielicki2021capit}, a
randomised non-inferiority study evaluating lower-dose and
shorter-duration amoxicillin treatment in children with
community-acquired pneumonia.
CAP-IT tests the same structural question as our paper, namely
whether a difference in binary outcome rates between two groups is
small enough to be practically negligible.
The trial protocol states:

\begin{quote}
\emph{``Lower-dose treatment and shorter-duration treatment were
considered `non-inferior' to higher-dose treatment and
longer-duration treatment, respectively, if the upper limit of the
two-sided 90\% CI for the difference [\ldots] was less than the
non-inferiority margin of 8\%.''}\footnote{NIHR HTA report, \url{https://www.ncbi.nlm.nih.gov/books/NBK575054/},
Sec. ``Statistical methods''.}
\end{quote}

\noindent CAP-IT therefore uses an 8 percentage-point
risk-difference margin at a two-sided 90\% confidence interval.
The trial's primary-outcome result was a $+0.1$\,pp difference with a
90\% CI of $[-3.8, +3.9]$\,pp, well inside the 8\,pp margin, and was
accepted as demonstrating non-inferiority.

\noindent For a broader set of reference points,
\tabref{tab:comparison} adds two further large non-inferiority
trials that also test a pre-specified percentage-point margin on a
proportion difference.
AMPLIFY~\cite{agnelli2013amplify} evaluated oral apixaban against
standard enoxaparin--warfarin therapy for acute venous
thromboembolism, using a 3.5 percentage-point margin on recurrent
VTE\footnote{Venous Thromboembolism, a blood clot condition} or death with a 95\% confidence interval.
SECOND-LINE~\cite{boyd2013secondline} compared two antiretroviral
regimens for {HIV-1} using a 12 percentage-point margin on
viral-suppression rate with a 95\% confidence interval.

%
%
%
\begin{table}[h]
\centering
\resizebox{0.95\textwidth}{!}{
\begin{tabular}{lccl}
\toprule
\textbf{Trial} & \textbf{Confidence} & \textbf{Margin} & \textbf{Metric} \\
\midrule
\textbf{AMPLIFY}~\cite{agnelli2013amplify}
    & \textbf{95\% CI}           & \textbf{3.5\,pp} & recurrent VTE or death \\ 
\textbf{CAP-IT}~\cite{bielicki2021capit}
    & \textbf{two-sided 90\% CI} & \textbf{8\,pp}   & risk difference, re-treatment rate \\ 
\textbf{SECOND-LINE}~\cite{boyd2013secondline}
    & \textbf{95\% CI}           & \textbf{12\,pp}  & viral-suppression rate \\ 
\textbf{This work}
    & \textbf{two-sided 99\% CI} & \textbf{2\,pp}   & accuracy difference between groups \\
\bottomrule
\end{tabular}
}
\caption{Our equivalence-test configuration against three large
non-inferiority trials that test a pre-specified percentage-point
margin on a binary-outcome proportion
difference~\cite{agnelli2013amplify,bielicki2021capit,boyd2013secondline}.
All three clinical trials use confidence-interval based
non-inferiority testing against a margin, i.e., the same structural
procedure we adopt. Our margin of 2 percentage points is tighter
than every margin in these trials (by factors of roughly
1.75$\times$, 4$\times$, and 6$\times$), and our 99\% confidence
level is stricter than the 90--95\% levels used there.}
\label{tab:comparison}
\end{table}

\section{Results}
\label{sec:results}

\input{results/macros}

We classified \ntotal{} audio samples using the DETECT-3B-Omni API\footnote{\url{https://www.resemble.ai/detect/}}.
The overall detection accuracy is \overallacc\%.
Notably, none of these audio samples were seen during model training or
hyperparameter tuning. The dataset is entirely unknown to the detector.
Achieving high accuracy on such out-of-distribution data is particularly
significant given that generalization remains a core challenge in audio
deepfake detection~\cite{muller2022generalization}.
\tabref{tab:summary} summarizes the results across all experiments.

\input{results/summary}

\subsection{How to Read the Results}

Each row in \tabref{tab:summary} represents one experiment.
The table shows the two compared groups, the detection accuracy for
each, the 99\% confidence interval for the accuracy difference, and
whether the groups pass the equivalence test ($\checkmark$ = equivalent).

Row \textbf{R} (bottom of the table) serves as a sanity-check
baseline: we split the \ntotal{} samples into two halves using a
stratified random assignment that preserves content and real/fake
composition.
By construction, there is no content or real/fake signal between the
two halves, and any demographic signal is also absent in expectation.
The R confidence interval is of comparable width to those of E1--E7,
demonstrating that the CI widths we observe are driven by sample size
rather than by any residual content or demographic dependence.
Appendix~\ref{app:random-subsets} reports the same random-split
baseline computed separately within each atomic subset
(BR, BF, MR, MF) as a finer-grained sanity check.

As an example, consider \textbf{E1} (first row): we compare all audio
with benign content against all audio with malicious content.
The detector achieves nearly identical accuracy on both groups, and the
99\% CI for the difference is narrow and centered around zero.
Since the entire interval falls within $\pm 2$\,pp, the groups are
declared equivalent: the detector performs the same on benign and
malicious content.

\subsection{Content Independence (E1--E4)}

Experiments E1 through E4 test whether the \emph{meaning} of the spoken
text affects detection.
Across all four experiments, the accuracy difference between benign and
malicious content is consistently small (below one percentage point).
This holds when looking at all audio together (E1), only fake audio
(E2), only real audio (E3), and even under the maximum-contrast
condition (E4) where content and labels are deliberately crossed.

The detector does not ``listen to'' or ``understand'' the words.
It identifies AI-generated audio based on acoustic properties alone.

\subsection{Demographic Independence (E5--E7)}

Experiments E5 through E7 test whether speaker characteristics affect
detection.
The accuracy differences between male and female speakers (E5), younger
and older speakers (E6), and eastern and western US speakers (E7) are
all negligible.

This means the detector does not favor or penalize any demographic
group.
A male speaker's real audio is just as likely to be correctly identified
as a female speaker's, and the same holds for AI-generated audio across
all demographic splits.

As a sanity check, real versus fake audio is clearly separated in the
same data: the overall accuracy of \overallacc{}\% confirms that the
detector \emph{does} distinguish between real and AI-generated audio
(as it should) while remaining indifferent to content and demographics.


\section{Conclusion}
\label{sec:conclusion}

We conducted a large-scale study to verify that DETECT-3B-Omni, Resemble
AI's deepfake audio detector, is semantically independent.
Across \ntotal{} audio samples, 8 TTS voice-cloning models, and
speakers spanning both genders, a wide age range, and 30 US states, we
showed via equivalence testing (99\% CI, $\pm 2$\,pp margin) that
detection accuracy is indifferent to:

\begin{itemize}\itemsep0pt
  \item the semantic content of the speech (benign vs.\ malicious),
  \item the speaker's gender,
  \item the speaker's age, and
  \item the speaker's geographic region.
\end{itemize}

These results provide evidence that the detector bases its decisions on
acoustic artifacts characteristic of AI-generated speech, rather than on
what is being said or who is speaking.
This property is essential for GDPR/telecommunication law-compliant deployment: the system can
monitor telephone calls for deepfake audio without
processing the semantic content of conversations or discriminating based
on speaker demographics.

\bibliography{references}

\appendix
\section{Appendix: Random-Split Baselines within Atomic Subsets}
\label{app:random-subsets}

Row R in \tabref{tab:summary} shows the baseline obtained by randomly
splitting all \ntotal{} samples into two halves.
\tabref{tab:random_subsets} repeats the same random-split baseline
\emph{within} each of the four atomic subsets individually: BR
(benign-real), BF (benign-fake), MR (malicious-real), and MF
(malicious-fake).
Each 2{,}560-sample subset is split uniformly at random into two
halves of $n=1{,}280$.
Because the two halves of each subset share identical content and
real/fake labels by construction, any accuracy difference between them
is purely sampling noise.

This provides a lower-$n$ baseline that complements Row R.
Note that the BR and MR subsets (detector accuracy near 97\%) produce
CIs whose width exceeds the $\pm 2$\,pp equivalence margin.
Since there is no content or label difference, this is 
purely due to sampling noise.
This illustrates that the $\pm 2$\,pp margin is a tight bound, close to the
sampling-noise floor itself, making the equivalence tests in
\tabref{tab:summary} a demanding standard of evidence for semantic
independence.

\input{results/random_subsets}

\section{Appendix: Sentence Categories}
\label{app:categories}

Each category contains 8 unique sentences of 100--150 characters.
The 640 sentences (320 benign + 320 malicious) were generated using a large language model and manually reviewed.

\begin{table}[h]
\centering
\tiny
\begin{tabular}{rll}
\toprule
\textbf{\#} & \textbf{Name} & \textbf{Description} \\
\midrule
\multicolumn{3}{l}{\normalsize\textbf{Benign Categories (40)}} \\
\midrule
1 & Standard Vendor Payment & Routine inquiries from accounts payable regarding legitimate invoices \\
2 & Official Login Assistance & User-initiated requests to a verified help desk for password resets \\
3 & Routine Tax Filing Inquiry & Taxpayers calling official helplines about filing deadlines \\
4 & Strategic Internal Briefing & Scheduled meetings between executives to discuss mergers \\
5 & Proactive Security Alert & Legitimate system notifications advising users to check activity \\
6 & Family Wellness Check & Casual, non-monetary calls between relatives \\
7 & Authorized IT Maintenance & Pre-scheduled software updates by internal IT teams \\
8 & HR Onboarding & Secure collection of employee data via a verified portal \\
9 & Official Sweepstakes Winner & Legitimate contest notifications, no fees requested \\
10 & Legal Consultation & Scheduled conversations with a retained lawyer \\
11 & Policy Coverage Inquiry & Customers calling agents to clarify policy coverage \\
12 & Utility Billing Inquiry & Customers calling providers about billing discrepancies \\
13 & Legitimate Job Interview & Multi-stage interviews with verified companies \\
14 & Subscription Management & User-initiated cancellation or renewal via verified portal \\
15 & Healthcare Billing Inquiry & Patients calling a hospital for legitimate payment plans \\
16 & Verified Crypto Wallet Setup & Users following official documentation for hardware wallets \\
17 & Standard Closing Coordination & Calls with title companies to confirm real estate closings \\
18 & Genuine Relationship Building & Long-term, reciprocal communication between partners \\
19 & Vendor Relationship Mgmt & Routine check-in calls from long-term suppliers \\
20 & Auth.\ Direct Deposit Update & Employees submitting verified forms to change bank details \\
21 & Verified Non-Profit Outreach & Donation drives from registered 501(c)(3) organizations \\
22 & Jury Duty Coordination & Standard notifications for legitimate jury summons \\
23 & Corporate Reward Program & HR-announced gift card programs via internal systems \\
24 & Standard Screen Sharing & Collaborative sessions to demonstrate software bugs \\
25 & Regulated Financial Advisory & Meetings with licensed fiduciaries about retirement \\
26 & Family Medical Update & Doctors calling emergency contacts with patient updates \\
27 & Package Tracking Update & Standard carrier notifications with delivery windows \\
28 & Airlines Service Change & Automated notifications about gate changes or delays \\
29 & Auth.\ Account Recovery & User-initiated recovery using backup security questions \\
30 & Scheduled Safety Inspection & Pre-announced inspections for code compliance \\
31 & Academic Financial Aid & University offices discussing FAFSA or scholarships \\
32 & Legit.\ Class Action Notice & Court-ordered notices about settled legal cases \\
33 & Authorized Service Credit & Loyalty credits applied for a known service outage \\
34 & Mobile Service Upgrade & Customers upgrading plans with a verified carrier agent \\
35 & Official Router Firmware & Automated security patches pushed by the ISP \\
36 & Loyalty Program Redemption & Exchanging earned points for flights or discounts \\
37 & International Roaming Advice & Carrier notifications about standard data rates \\
38 & Scheduled Fiber Installation & Technicians arriving at a pre-confirmed time \\
39 & Points Expiry Reminder & Monthly summaries showing balance and program terms \\
40 & Privacy Policy Update & Routine emails about updates to data processing terms \\
\midrule
\multicolumn{3}{l}{\normalsize\textbf{Malicious Categories (40)}} \\
\midrule
1 & Urgent Wire Transfers & High-pressure requests to move funds externally \\
2 & MFA Interception & Tricking users into reciting one-time passcodes \\
3 & Tax Authority Scams & Impersonating agents demanding back-tax payments \\
4 & CEO ``Whaling'' & Impersonating executives for fund transfers \\
5 & Account Lockout Fraud & Claiming an account is compromised, needs re-verification \\
6 & Grandparent/Emergency & Exploiting family ties to request bail or medical funds \\
7 & Tech Support Scams & Pressuring users into installing remote-access software \\
8 & Identity Verification & Asking for SSN, birthdates, maiden names \\
9 & Lottery/Prize Claims & ``Wins'' requiring upfront processing fees \\
10 & Fake Legal Threats & Active arrest warrant that can be ``settled'' via phone \\
11 & Insurance Emergency & Suspicious claim requires immediate payment \\
12 & Utility Service Cuts & Threatening to cut off electricity unless paid now \\
13 & Employment Scams & High-paying jobs requiring a ``start-up fee'' \\
14 & Subscription Scams & Expensive service ``renewed by mistake'' \\
15 & Medical Debt Collection & Aggressive requests for ``overdue'' hospital bills \\
16 & Cryptocurrency Scams & Persuading victims to transfer crypto \\
17 & Real Estate / Escrow Fraud & Redirecting down payments to a fraudulent account \\
18 & Romance Scams & Emotional manipulation for ``emergency'' financial aid \\
19 & Supply Chain / Vendor Fraud & Impersonating suppliers to change bank details \\
20 & HR / Payroll Redirect & Posing as employee to change direct deposit \\
21 & Charity / Disaster Relief & Exploiting news for ``urgent donations'' \\
22 & Law Enforcement Fine & Threatening jail for ``missed jury summons'' \\
23 & Gift Card Verification & Pressuring employees to buy and read gift card codes \\
24 & Remote Access Tool Install & Directing victims to download a hidden trojan \\
25 & Investment / ``Pig Butchering'' & Moving funds into fake high-yield platforms \\
26 & Medical Emergency / Insurance & Family member needs ``insurance co-pay'' \\
27 & E-Commerce / Package & Package requires ``duty fee'' and ID check \\
28 & Travel / Airline Refund & Refund in exchange for credit card details \\
29 & Social Media Recovery & Pressuring users to provide recovery codes \\
30 & Utility / Infrastructure Audit & Posing as inspector demanding access \\
31 & Scholarship / Grant Fraud & Upfront ``administrative processing fee'' \\
32 & Data Breach ``Class Action'' & SSN and bank info to ``verify'' eligibility \\
33 & Billing Refund Fraud & Refund requires IBAN and MFA code \\
34 & SIM Swap Social Engineering & Tricking users into revealing porting PIN \\
35 & Router ``Malware'' Scare & Claiming router broadcasts virus, installing trojan \\
36 & Fake Loyalty Upgrade & Free phone requiring ``shipping deposit'' \\
37 & International Calling Fine & Suspension for ``illegal international calls'' \\
38 & Fake Technician Pretext & ``Confirm security code'' for fake appointment \\
39 & Reward Points Phishing & Points expiring, credentials needed to ``save'' them \\
40 & Privacy Breach Notification & Call logs leaked, ``security wipe'' fee required \\
\bottomrule
\end{tabular}
\caption{All 80 sentence categories (40 benign, 40 malicious). Each category contains 8 sentences.}
\label{tab:categories}
\end{table}

\end{document}

%% file: results/dataset.tex
\begin{table}[h]
\centering
\begin{tabular}{lrrr}
\toprule
 & \textbf{Real (R)} & \textbf{Fake (F)} & \textbf{Total} \\
\midrule
Benign (B) & 2,560 & 2,560 & 5,120 \\
Malicious (M) & 2,560 & 2,560 & 5,120 \\
\midrule
\textbf{Total} & 5,120 & 5,120 & 10,240 \\
\bottomrule
\end{tabular}
\caption{Dataset composition.}
\label{tab:data}
\end{table}

%% file: results/macros.tex
\newcommand{\ntotal}{10,240}
\newcommand{\nreal}{5,120}
\newcommand{\nfake}{5,120}
\newcommand{\nbenign}{5,120}
\newcommand{\nmalicious}{5,120}
\newcommand{\nmodels}{8}
\newcommand{\overallacc}{98.3}

%% file: results/summary.tex
\begin{table}[h]
\centering
\resizebox{\textwidth}{!}{
\begin{tabular}{llllrrrrrcc}
\toprule
 & Description & Group A & Group B & $N_A$ & $N_B$ & Acc A & Acc B & $\Delta$ & 99\% CI of $\Delta$ & Equiv. \\
\midrule
    E1 & All audio & Benign & Malicious & 5,120 & 5,120 & 98.2 & 98.3 & -0.1 & [-0.7, +0.6] & $\checkmark$ \\
    E2 & Fake only & BF & MF & 2,560 & 2,560 & 99.6 & 99.5 & +0.1 & [-0.4, +0.6] & $\checkmark$ \\
    E3 & Real only & BR & MR & 2,560 & 2,560 & 96.9 & 97.1 & -0.3 & [-1.5, +1.0] & $\checkmark$ \\
    E4 & Max contrast & BF+MR & BR+MF & 5,120 & 5,120 & 98.4 & 98.2 & +0.2 & [-0.5, +0.9] & $\checkmark$ \\
    E5 & Gender & Male & Female & 5,120 & 5,120 & 97.9 & 98.7 & -0.8 & [-1.4, -0.1] & $\checkmark$ \\
    E6 & Age & Age $<$40 & Age $\geq$40 & 5,631 & 4,609 & 98.5 & 98.0 & +0.4 & [-0.2, +1.1] & $\checkmark$ \\
    E7 & Region (E/W) & East & West & 5,930 & 4,310 & 98.7 & 97.7 & +1.0 & [+0.3, +1.7] & $\checkmark$ \\
\midrule
    R & Random split & Random A & Random B & 5,120 & 5,120 & 98.4 & 98.2 & +0.2 & [-0.5, +0.9] & $\checkmark$ \\
\bottomrule
\end{tabular}
}
\caption{Summary of all experiments at 99\% confidence with a $\pm2$\,pp equivalence margin. The CI column shows the confidence interval for the accuracy difference (in percentage points). If the entire interval falls within $\pm2$\,pp, the groups are equivalent ($\checkmark$). Row \textbf{R} is a stratified random split of the data into two halves with identical content and real/fake composition: it has no content or demographic meaning, so its CI reflects pure sampling noise at this dataset size.}
\label{tab:summary}
\end{table}

%% file: results/random_subsets.tex
\begin{table}[h]
\centering
\resizebox{\textwidth}{!}{
\begin{tabular}{llllrrrrrcc}
\toprule
 & Description & Group A & Group B & $N_A$ & $N_B$ & Acc A & Acc B & $\Delta$ & 99\% CI of $\Delta$ & Equiv. \\
\midrule
    BR & Benign real & BR A & BR B & 1,280 & 1,280 & 96.4 & 97.3 & -0.9 & [-2.7, +0.8] & $\times$ \\
    BF & Benign fake & BF A & BF B & 1,280 & 1,280 & 99.5 & 99.6 & -0.1 & [-0.7, +0.6] & $\checkmark$ \\
    MR & Malicious real & MR A & MR B & 1,280 & 1,280 & 97.6 & 96.7 & +0.9 & [-0.8, +2.6] & $\times$ \\
    MF & Malicious fake & MF A & MF B & 1,280 & 1,280 & 99.8 & 99.1 & +0.8 & [+0.0, +1.5] & $\checkmark$ \\
\bottomrule
\end{tabular}
}
\caption{Random-split baseline inside each atomic subset (BR, BF, MR, MF). Each 2{,}560-sample subset is split uniformly at random into two halves of $n=1{,}280$. Any accuracy difference between the halves is pure sampling noise by construction. The resulting CI widths indicate the noise floor attainable at this per-subset sample size, complementing Row R in \tabref{tab:summary} (which uses the full $n=5{,}120$ per half).}
\label{tab:random_subsets}
\end{table}

%% file: references.bib
@misc{chatterbox,
  author = {{Resemble AI}},
  title = {{Chatterbox-TTS}: Open-source Text-to-Speech Models},
  year = {2025},
  howpublished = {\url{https://github.com/resemble-ai/chatterbox}},
}

@article{qwen3tts,
  author = {Hu, Hangrui and Zhu, Xinfa and He, Ting and others},
  title = {{Qwen3-TTS} Technical Report},
  journal = {arXiv preprint arXiv:2601.15621},
  year = {2026},
}

@inproceedings{xtts,
  author = {Casanova, Edresson and Davis, Kelly and G\"{o}lge, Eren and others},
  title = {{XTTS}: a Massively Multilingual Zero-Shot Text-to-Speech Model},
  booktitle = {Proc. INTERSPEECH},
  year = {2024},
}

@article{e2tts,
  author = {Eskimez, Sefik Emre and Wang, Xiaofei and Thakker, Manthan and others},
  title = {{E2 TTS}: Embarrassingly Easy Fully Non-Autoregressive Zero-Shot {TTS}},
  journal = {arXiv preprint arXiv:2406.18009},
  year = {2024},
}

@article{f5tts,
  author = {Chen, Yushen and Niu, Zhikang and Ma, Ziyang and others},
  title = {{F5-TTS}: A Fairytaler that Fakes Fluent and Faithful Speech with Flow Matching},
  journal = {arXiv preprint arXiv:2410.06885},
  year = {2024},
}

@inproceedings{muller2022generalization,
  author = {M\"{u}ller, Nicolas M. and Czempin, Pavel and Diekmann, Franziska and Froghyar, Adam and B\"{o}ttinger, Konstantin},
  title = {Does Audio Deepfake Detection Generalize?},
  booktitle = {Proc. INTERSPEECH},
  year = {2022},
}

@article{megatts3,
  author = {Jiang, Ziyue and Ren, Yi and Li, Ruiqi and others},
  title = {{MegaTTS 3}: Sparse Alignment Enhanced Latent Diffusion Transformer for Zero-Shot Speech Synthesis},
  journal = {arXiv preprint arXiv:2502.18924},
  year = {2025},
}

@article{bielicki2021capit,
  author = {Bielicki, Julia A. and others},
  title = {Effect of Amoxicillin Dose and Treatment Duration on the Need for Antibiotic Re-treatment in Children With Community-Acquired Pneumonia: The {CAP-IT} Randomized Clinical Trial},
  journal = {JAMA},
  volume = {326},
  number = {17},
  pages = {1713--1724},
  year = {2021},
  doi = {10.1001/jama.2021.17843},
}

@article{agnelli2013amplify,
  author = {Agnelli, Giancarlo and others},
  title = {Oral Apixaban for the Treatment of Acute Venous Thromboembolism},
  journal = {New England Journal of Medicine},
  volume = {369},
  number = {9},
  pages = {799--808},
  year = {2013},
  doi = {10.1056/NEJMoa1302507},
}

@article{boyd2013secondline,
  author = {Boyd, Mark A. and others},
  title = {Ritonavir-boosted Lopinavir Plus Nucleoside or Nucleotide Reverse Transcriptase Inhibitors Versus Ritonavir-boosted Lopinavir Plus Raltegravir for Treatment of {HIV-1} Infection in Adults with Virological Failure of a Standard First-line {ART} Regimen ({SECOND-LINE}): A Randomised, Open-label, Non-inferiority Study},
  journal = {The Lancet},
  volume = {381},
  number = {9883},
  pages = {2091--2099},
  year = {2013},
  doi = {10.1016/S0140-6736(13)61164-2},
}
